\documentclass[twocolumn]{svjour3}          
\smartqed  
\usepackage{graphicx}
\usepackage{mathptmx}      
\usepackage{authblk}
\usepackage{amsmath, amsfonts}
\usepackage{bm} 
\usepackage{units}
\usepackage{physics}
\usepackage{import}
\usepackage{hyphenat}
\usepackage[utf8]{inputenc}
\usepackage[colorlinks=true,allcolors=blue]{hyperref}
\usepackage{subcaption}
\usepackage[numbers]{natbib}

\renewcommand{\curl}{\,\mathrm{curl}\,}
\renewcommand{\div}{\,\mathrm{div}\,}
\renewcommand{\grad}{\,\mathrm{grad}\,}
\newcommand{\curll}{\mathrm{curl}\,}
\newcommand{\divv}{\mathrm{div}\,}
\newcommand{\gradd}{\mathrm{grad}\,}
\newcommand{\iu}{j} 
\journalname{}
\begin{document}

\title{Broadband Finite-Element Impedance Computation for Parasitic Extraction} 



%

\author{J. Stysch  \and
        A. Klaedtke \and
        H. De Gersem
}

\institute{J. Stysch \and A. Klaedtke \at
              Robert Bosch GmbH \\
              71272 Renningen, Germany\\
              \email{jonathan.stysch@de.bosch.com}           
           \and
          H. De Gersem \at
              Institute for Accelerator Science and Electromagnetic Fields\\ Technical University of Darmstadt
}

\date{Received: date / Accepted: date}

\maketitle

\begin{abstract}
Parasitic extraction is a powerful tool in the design process of electromechanical devices, specifically as part of workflows that check electromagnetic compatibility. 
A novel scheme to extract impedances from CAD device models, suitable for a finite element implementation, is derived from Maxwell's equations in differential form. 
It provides a foundation for parasitic extraction across a broad frequency range and is able to handle inhomogeneous permittivities and permeabilities, making it more flexible than existing integral equation approaches. 
The approach allows for the automatic treatment of multi-port models of arbitrary conductor geometry without requiring any significant manual user interaction. 
This is achieved by computing a connecting source current density that supplies current to the model's terminals, whatever their location in the model, subsequently using this current density to compute the electric field, and finally calculating the impedance via a scalar potential. 
A mandatory low-frequency stabilization scheme is outlined, ensuring a stable evaluation of the model at low frequencies as well. 
Two quasistatic approximations and the special case of perfect electric conductors are treated theoretically. 
The magnetoquasistatic approximation is validated against an analytical model in a numerical experiment. 
Moreover, the intrinsic capability of the method to treat inhomogeneous permittivities and permeabilities is demonstrated with a simple capacitor-coil model including dielectric insulation and magnetic core materials.
\keywords{Parasitic effects \and Finite element method \and Quasistatics \and Electromagnetic compatibility}
\end{abstract}

\section{Introduction}
\label{sec:intro}
Increasing switching frequencies in power electronics, miniaturization and stricter electromagnetic compatibility (EMC) regulations pose a big challenge for the development of electromechanical devices. Numerical simulations used in design and optimization workflows are commonplace. Simulations do not only allow to assess a design before prototyping but also to analyze the electromagnetic behavior of models that are difficult to access in measurements due to, e.g., very compact dimensions. An important task in EMC analysis is to quantify the influence of the non\hyp{}functional elements like interconnects or chassis of a device under test (DUT) on its intended functionality. This is typically done by extracting \emph{parasitic components} from a model of the DUT in order to complement circuit models that contain the functional elements. Parasitic components can either be simply frequency\hyp{}independent resistances, inductances, and capacitances \cite{ruehli_equivalent_1974}  or frequency\hyp{}dependent (reduced order) models of transfer functions like the impedance, admittance and scattering matricies $\mathbf{Z}$, $\mathbf{Y}$, and $\mathbf{S}$, respectively \cite{kamon_automatic_1998}.

The partial element electric circuit (PEEC) method \cite{ruehli_equivalent_1974} and further developments thereof \cite{kamon_automatic_1998,kamon_fasthenry:_1994} are classical solution techniques in this context. This class of methods is based on Maxwell's equations in integral form, using Green's functions in the solution process, which forbids any straightforward treatment of spatially inhomogeneous permittivities $\varepsilon$ or permeabilities $\mu$. In their standard form the methods of this class require a spatial discretization into cuboid volume elements, which can be impractical to represent the complex geometries encountered in industrial applications. Generalizing these methods with less restrictive volume elements may be cumbersome \cite{ruehli_nonorthogonal_2003}.

These disadvantages are avoided with an approach based on the finite element (FE) method, which allows for a highly flexible spatial discretization (commonly using tetrahedral elements), and an inherent treatment of inhomogeneous material parameters. In \cite{traub_generation_2012} an FE-based method for impedance computation was introduced in a rudimentary fashion, lacking an appropriate theoretical treatment, and considering only the special case of homogeneous material parameters and lossless conductors in the Darwin approximation. Nevertheless, a successful application of this method served as the foundation of a geometrical sensitivity analysis in \cite{schuhmacher_adjoint_2018} and confirmed its potential. The aim of this paper is to provide a solid theoretical foundation for a more general impedance computation method based on the approach outlined in \cite{traub_generation_2012}. The such computed impedance matrix $\mathbf{Z}$ may subsequently be used to extract parasitic lumped elements or to compute the $\mathbf{Y}$ and $\mathbf{S}$ transfer functions.

Section \ref{sec:voltage} describes how the impedance of a conductor segment can be calculated from field quantities by introducing a reduced voltage, which can ultimately be calculated from the electric scalar potential $\phi$. The main section \ref{sec:bvp_general} provides the derivation of the field-theoretical model determining $\phi$ via the E-field formulation of Maxwell's equations. A connecting excitation current density is introduced, and any unwanted inductive influence of this excitation current is de-embedded from the result for $\phi$ by a compensation term. Two quasistatic approximations are provided, and the special case of lossless conductors for a high-frequency approximation of the inductance is discussed. Finally, a necessary low-frequency stabilization scheme is outlined. In section \ref{sec:experiments} numeric results of the proposed method are compared to analytical impedance and inductance values of a wire, and the capability of the method to handle inhomogeneous permittivities $\varepsilon$ and permeabilities $\mu$ is demonstrated with a model featuring dielectric insulation and a magnetic core.  A brief summary concludes this paper in section \ref{sec:conclusion}.
\section{Voltage Calculation}
\label{sec:voltage}
To reconcile the path-dependent voltage concept of electromagnetic field theory with the path\hyp{}independent voltage concept of electrical circuits poses a challenge. In this section, we consider the example of a thin wire in a plane as discussed in \cite[Chapter~14.16]{jordan_electromagnetic_1968} and heuristically generalize the result to arbitrary three\hyp{}dimensional conductors.

\begin{figure}
	\centering
	\includegraphics[scale=1]{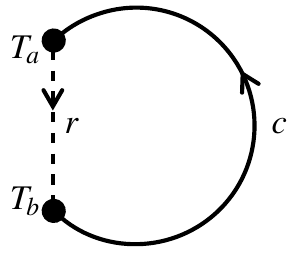}
	\caption{\label{fig:loop} Wire $c$ with terminals $T_a$ and $T_b$ and return path $r$.}
\end{figure}

Fig.~\ref{fig:loop} displays a path $c$ along a wire, which ends in the terminals $T_a$ and $T_b$. A return path $r$ connecting the two terminals forms a closed loop with $c$, that encloses the surface $S_0$. Integrating Faraday's law in frequency domain over $S_0$ and applying Stokes' theorem yields
\begin{equation}
\oint\limits_{\partial S_0} \bm{E}\cdot d\bm{l} = \int\limits_{r}\bm{E}\cdot d\bm{l} + \int\limits_{c}\bm{E}\cdot d\bm{l} = - \iu\omega \int\limits_{S_0}\bm{B}\cdot d\bm{S},
\end{equation} 
with $\bm{E}$ and $\bm{B}$ denoting the electric field strength and the magnetic flux density, respectively.
The voltage $V$ that has to be applied at the terminals in order to move charges from $T_a$ to $T_b$ against the self-induced electric field $\bm{E}$ must be the negative of the integral over $r$, which yields
\begin{equation}\label{voltage_1}
V = -\int\limits_{r}\bm{E}\cdot d\bm{l} = \int\limits_{c}\bm{E}\cdot d\bm{l} +\iu\omega \int\limits_{S_0}\bm{B}\cdot d\bm{S}.
\end{equation}

This is the voltage an ideal voltmeter would measure between the terminals. The first integral on the right-hand side captures the voltage drop due to the internal impedance of the wire (i.e., for a one-dimensional wire, the ohmic resistance) while the second integral is associated with the external reactance of the loop \cite{jordan_electromagnetic_1968}. For frequencies sufficiently below the first resonance of the system, the surface integral over $\bm{B}$ divided by the current $I$ causing the magnetic field is the inductance of the loop,
\begin{equation}
L_\mathrm{loop} = \frac{1}{I} \int\limits_{S_0}\bm{B}\cdot d\bm{S}. 
\end{equation}
However, for the purpose of extracting parasitics to be used in circuit simulation, we are not interested in the inductive response of the closed loop but only in the part of the response due to the wire. In an experiment, the return current path $r$ closing the loop would correspond to either a current source or a voltmeter connected to the terminals. We therefore want to exclude the part of the inductive response due to the path $r$ between the terminals and, moreover, avoid having to specify the path closing the loop altogether.

This is achieved by employing the concept of partial inductance (see, e.g., \cite{paul_inductance:_2010}): The partial inductances $L_c$ and $L_{r}$ of the respective segments of the loop can be calculated by expressing the magnetic field with the vector potential $\bm{A}$ as $\bm{B} = \curl \bm{A}$ and applying Stokes' theorem:
\begin{equation}
L_\mathrm{loop} = \frac{1}{I} \oint\limits_{\partial S_0}\bm{A}\cdot d\bm{l} =  \frac{1}{I} \int\limits_{c}\bm{A}\cdot d\bm{l} +  \frac{1}{I} \int\limits_{r}\bm{A}\cdot d\bm{l} =:  L_c + L_{r} .
\end{equation}
The values of these partial inductances depend on the gauge condition chosen for the magnetic vector potential $\bm{A}$ and electric scalar potential $\phi$.

Subtracting the inductive contribution $\iu\omega IL_{r}$ of the path between the terminals from the voltage of \eqref{voltage_1} allows to define a reduced ``conductor voltage'' $V_\mathrm{c}$ that contains only the resistive and inductive response of the wire $c$,
\begin{equation}\label{eqn:voltage_reduced}
V_\mathrm{c} := V - I\iu\omega L_{r} =  \int\limits_{c}\bm{E}\cdot d\bm{l} + \iu\omega \int\limits_{c}\bm{A}\cdot d\bm{l}.
\end{equation}
Expressing the electric field in \eqref{eqn:voltage_reduced} as
\begin{equation}\label{eqn:E_potential}
	\bm{E} = -\grad\phi -\iu\omega \bm{A}
\end{equation}
yields the simplification
\begin{equation}\label{eqn:voltage_reduced2}
	V_\mathrm{c} = -\int\limits_{c}\gradd \phi \cdot d\bm{l} = \phi(T_b)-\phi(T_a) .
\end{equation}
Thus, the reduced voltage is in fact just the potential difference of the scalar potential $\phi$ between the two terminals and thereby path independent. As it incorporates a partial inductance, $V_\mathrm{c}$ formally depends on the gauge condition of potentials $\phi$ and $\bm{A}$.

The path independence of $V_\mathrm{c}$ facilitates a heuristic generalization to three-dimensional conductors, where the terminals are surfaces instead of points. The reduced voltage can be calculated analogously to the thin\hyp{}wire case of  \eqref{eqn:voltage_reduced2} by simply averaging the potential over the respective surfaces,
\begin{equation}\label{eqn:voltage_reduced_3d}
V_\mathrm{c} := \frac{1}{A(T_b)}\int\limits_{T_b} \phi\, dS - \frac{1}{A(T_a)}\int\limits_{T_a} \phi\, dS.
\end{equation}
Here, $A(T_a)$ and $A(T_b)$ denote the surface areas of the two terminal surfaces.

In the general case, there may be several conductors that may each have multiple terminals. To calculate an $N\times N$ impedance matrix $\mathbf{Z}$, a topology of $N$ branches connecting the terminals must be provided. A series of $N$ numerical experiments can then be conducted, in each of which the current $I_0$ flows through one of the branches $j$. The element $Z_{ij}$ of the inductance matrix is given through the voltage $V_{ij}$ between the two terminals of branch $i$ (which is calculated using \eqref{eqn:voltage_reduced_3d}),
\begin{equation}\label{eqn:impedance}
	Z_{ij} = \frac{V_{ij}}{I_0}.
\end{equation}

\section{Field-Theoretical Model for Impedance Computation}\label{sec:bvp_general}
\subsection{Formulation of the Fundamental Differential Equations}
The previous section showed how the impedance matrix $\mathbf{Z}$ is calculated from the electric scalar potential $\phi$. 
For a system excited by a given source current density $\bm{J}_\mathrm{s}$, two possible strategies to calculate $\phi$  are available:\\
The first option is to solve Maxwell's equations directly in a potential formulation for the electric scalar potential $\phi$ and the magnetic vector potential $\bm{A}$. The second option is to first calculate $\bm{E}$ by solving the `E-field formulation' (see, e.g, \cite{monk_finite_2003}), and subsequently calculate $\phi$ in a second step using \eqref{eqn:E_potential} and a gauge condition. 
It is more advantageous to use the latter ``$\bm{E}$ approach'' for several reasons: The two fields $\bm{E}$ and $\phi$ can be calculated in sequence (except for the Darwin approximation case discussed in section \ref{sec:darwin}), thereby avoiding a computationally more expensive coupled boundary value problem (BVP), which occurs in the ``$\bm{A}$-$\phi$ approach''. Furthermore, the $\bm{E}$ approach allows for an easy treatment of conductors modeled as perfect electric conductors (PECs), which is an important special case discussed in section \ref{sec:pec}. Finally, the use of the E-field formulation allows for an elegant stabilization of the low-frequency instability inherent to all FE formulations of Maxwell's equations in frequency domain, which is discussed in section \ref{sec:lfs}.

The E-field formulation can be derived combining Am\-père's law and Faraday's law. To provide a complete boundary value problem (BVP) that forms the basis of a FE solution, the boundary $\partial \Omega$ of the computational domain $\Omega$ is assumed to be the union of an electric and a magnetic boundary, $\Gamma_\mathrm{el}$ and $\Gamma_\mathrm{mag}$, on which electric (Dirichlet) and magnetic (Neumann) boundary conditions, respectively, are to be applied:
\begin{equation}\label{eqn:outer_boundary}
\partial \Omega = \Gamma_\mathrm{el} \cup \Gamma_\mathrm{mag}.
\end{equation}
Commonly either $\Gamma_\mathrm{el}$ or $\Gamma_\mathrm{mag}$ is empty, such that $\partial \Omega$ is entirely electric or magnetic. The BVP of the E-field formulation thereby reads,
\begin{subequations} \label{eqn:E_field_form}
	\begin{align}
	\curll \nu_\mathrm{r} \curl \bm{E} +  \iu\omega\mu_0\sigma\bm{E} &- \frac{\omega^2}{c^2}\varepsilon_\mathrm{r}\bm{E} &&\notag\\
	&= -\iu\omega \mu_0\bm{J}_\mathrm{s} && \mathrm{in} \; \Omega,\label{eqn:E_field_form_a}\\
	\bm{n}\times\bm{E} &= 0  && \mathrm{on} \; \Gamma_\mathrm{el},\\
	\bm{n}\cdot\varepsilon_\mathrm{r}\bm{E} &= 0  && \mathrm{on} \; \Gamma_\mathrm{mag}. \label{eqn:mbc}
	\end{align}
\end{subequations}
Here, $\mu_0$ is the permeability of the vacuum, and $c$ the speed of light, while $\sigma$, $\varepsilon_\mathrm{r}$ and $\nu_\mathrm{r}$ are the spatially dependent conductivity, relative permittivity and relative reluctivity, respectively, and $\bm{n}$ is the normal vector on $\partial \Omega$. The magnetic boundary condition for $\bm{E}$ \eqref{eqn:mbc} can be derived from the magnetic boundary condition for the magnetic field strength $\bm{H}$,  $\bm{n}\times\bm{H} = 0$, using Am\-père's law and demanding $\bm{n} \cdot (\sigma\bm{E} + \bm{J}_\mathrm{s}) =0$ on $\Gamma_\mathrm{mag}$.

The partial differential equation (PDE) to determine $\phi$ is found by using \eqref{eqn:E_potential} to eliminate the vector potential $\bm{A}$ from a gauge condition. Here, the Lorenz gauge \cite{nisbet_electromagnetic_1957},
\begin{equation}\label{eqn:gauge}
	\divv \varepsilon_\mathrm{r} \bm{A} + \frac{\iu\omega}{c^2}\phi = 0,
\end{equation}
is chosen since it enables the calculation of the inductive compensation term introduced in section \ref{sec:compensation}. Together \eqref{eqn:E_potential} and \eqref{eqn:gauge} yield the PDE
\begin{equation}\label{eqn:phi_1}
-\divv\varepsilon_\mathrm{r}\grad\phi -\frac{\omega^2}{c^2}\phi= \div\varepsilon_\mathrm{r} \bm{E}
\end{equation}
The full BVP to compute the electric scalar potential is given in section \ref{sec:compensation}, after modeling the source current density $\bm{J}_\mathrm{s}$ and introducing the compensation term.

\subsection{Modeling the Connecting Source Current Density}
The remaining task is to identify a suitable source current density $\bm{J}_\mathrm{s}$ modeling a current source connected to two of the terminals of the DUT. The source current density $\bm{J}_\mathrm{s}$ should inject a constant current $I_0$ at the terminal surface $T_b$ and extract the same current again at $T_a$, such that it flows along the integration path $c$ in Fig.~\ref{fig:loop}. Inside the conductor, this current $I_0$ is transported between the terminals by a conduction current density $\bm{J}_\mathrm{c}$, which unlike $\bm{J}_\mathrm{s}$ directly couples to $\bm{E}$ via Ohm's law, $\bm{J}_\mathrm{c} = \sigma \bm{E}$, and thereby captures the resistive response to the enforced current flow. For the current densities $\bm{J}_a$ and $\bm{J}_b$ on the terminal surfaces homogeneous distributions are chosen,
\begin{equation}
\bm{J}_a = \hat{\bm{n}}_a\frac{I_0}{A(T_a)}\;\;\mathrm{and}\;\; \bm{J}_b = -\hat{\bm{n}}_b\frac{I_0}{A(T_b)},
\end{equation}
with $\hat{\bm{n}}_i$ denoting the unit normal vector pointing out of the conductor, and $A(T_i)$ again the area of the respective terminals.
The associated divergence of $\bm{J}_\mathrm{s}$ must hence be given by
\begin{equation} \label{eqn:div_J}
-\divv\bm{J}_\mathrm{c} = \divv \bm{J}_\mathrm{s} =
|\bm{J}_a|\,\delta_a(\bm{r})
-|\bm{J}_b|\,\delta_b(\bm{r})	
\end{equation}
where the delta distribution $\delta_i(\bm{r})$ at terminal $T_i$ is defined by 
\begin{equation}
	\int\limits_V \delta_i(\bm{r}) f(\bm{r}) dV = \int\limits_{T_i} f(\bm{r}) dS\quad \forall f(\bm{r}) : \mathbb{R}^3 \rightarrow \mathbb{C}.
\end{equation}
The expression \eqref{eqn:div_J} describes the situation that at the terminals the source current density $\bm{J}_\mathrm{s}$ takes over the task of transporting the current $I_0$ from the conduction current density $\bm{J}_\mathrm{c}$.

Our method to calculate $\bm{J}_\mathrm{s}$ has to be so general that it can easily produce a connecting $\bm{J}_\mathrm{s}$ independently of how the terminal surfaces are positioned in relation to e.g. the outer boundary $\partial\Omega$ or any of the conducting areas of the DUT. Generally, an expression for the divergence of the vector field $\bm{J}_\mathrm{s}$ is not sufficient to determine $\bm{J}_\mathrm{s}$. However, choosing a gradient field ansatz,
\begin{equation}\label{eqn:grad_g}
\bm{J}_\mathrm{s} = - \tilde{\sigma}\gradd \xi,
\end{equation}
enables a computation of $\bm{J}_\mathrm{s}$ from its divergence given in \eqref{eqn:div_J} without further specifying the path of the source current. The fictitious conductivity $\tilde{\sigma}$ is a parameter of the algorithm and is here chosen to be constant in the whole domain $\Omega$. The BVP determining the underlying potential $\xi$ (and therefore $\bm{J}_\mathrm{s}$) is equivalent to the standard stationary current problem,
\begin{subequations}\label{eqn:xi}
	\begin{align}
	-\divv \tilde{\sigma} \grad \xi  &=  \divv\bm{J}_\mathrm{s}  & & \mathrm{in} \; \Omega, \label{eqn:xi1} \\
	\xi &= \mathrm{const.} & & \mathrm{on} \; \Gamma_\mathrm{el},\label{eqn:xi2} \\
	\bm{n}\cdot\!\grad \xi &= 0 & & \mathrm{on} \; \Gamma_\mathrm{mag}.\label{eqn:xi3}
	\end{align}
\end{subequations}
The boundary conditions \eqref{eqn:xi2} and \eqref{eqn:xi3} are chosen in this way for consistency with the BVP \eqref{eqn:E_field_form} determining $\bm{E}$.

\subsection{Compensating the Inductive Influence of the Source Current Density}
\label{sec:compensation}
The gradient-field source current density proposed in the previous subsection does not model an ideal current source in one respect: Ideal current sources must not influence the DUT inductively. In section \ref{sec:voltage}, a reduced voltage $V_\mathrm{c}$ was defined in which the part of the inductive response related to the return path of the current was eliminated. In addition, the electromagnetic fields causing $V_\mathrm{c}$ should not capture any inductive influence of the source current $\bm{J}_\mathrm{s}$. The electric field calculated with \eqref{eqn:E_field_form} includes this unwanted influence of the source current, since any source current density that has a component parallel to the DUT at a finite distance must be expected to have a direct inductive influence on the fields in the DUT (such parallel components are unavoidable for three-dimensional conductors of arbitrary shape).

It is, however, possible to quantify and eliminate (and thereby de-embed) the contribution of this unwanted direct influence of the source current density in the calculation of the scalar potential with \eqref{eqn:phi_1}. To this end, the ``total'' electric field $\bm{E}$ of \eqref{eqn:E_field_form}, incorporating inductive effects related to both the conductor currents and the source current, is expressed as the difference of the compensated field $\bm{E}_\mathrm{c}$ capturing only the influence of the conductors and the field $\bm{E}_\mathrm{s}$ capturing the counteractive inductive influence of the source current density,
\begin{equation}
\bm{E} = \bm{E}_\mathrm{c} - \bm{E}_\mathrm{s}.
\end{equation}
The scalar potential to be calculated with \eqref{eqn:phi_1} must be the compensated potential $\phi_\mathrm{c}$ excluding the unwanted inductive influence of $\bm{J}_\mathrm{s}$,
\begin{equation}\label{eqn:phi_2}
\begin{split}
-\divv\varepsilon_\mathrm{r}\grad\phi_\mathrm{c} -\frac{\omega^2}{c^2}\phi_\mathrm{c} &= \div\varepsilon_\mathrm{r} \bm{E}_\mathrm{c}\\ &= \div\varepsilon_\mathrm{r} \bm{E} + \div\varepsilon_\mathrm{r} \bm{E}_\mathrm{s} .
\end{split}
\end{equation}
The compensation term $\div\varepsilon_\mathrm{r} \bm{E}_\mathrm{s}$ can be determined employing the potential formulation of Maxwell's equations in Lorenz gauge,
\begin{subequations}\label{eqn:potential_form}
	\begin{align}
	\curll \nu_\mathrm{r} \curl \bm{A} - \varepsilon_\mathrm{r} \grad \div \varepsilon_\mathrm{r} \bm{A} - \frac{\omega^2}{c^2}\varepsilon_\mathrm{r}\bm{A}  =  \mu_0\bm{J}, \label{eqn:potential_form_1}\\
	-\divv \varepsilon_\mathrm{r} \grad \phi - \frac{\omega^2}{c^2}\phi = \frac{1}{\varepsilon_0}\rho .
	\end{align}
\end{subequations}
This formulation of Maxwell's equations enables the independent calculation of the vector potential $\bm{A}$ from the current density $\bm{J}$ and the scalar potential $\phi$ from the charge density $\rho$ in the case that all currents in a given system are source currents, $\bm{J} = \bm{J}_\mathrm{s}$, that do not couple to the electric field via Ohm's law.

\begin{figure}
	\centering
	\includegraphics[scale=1]{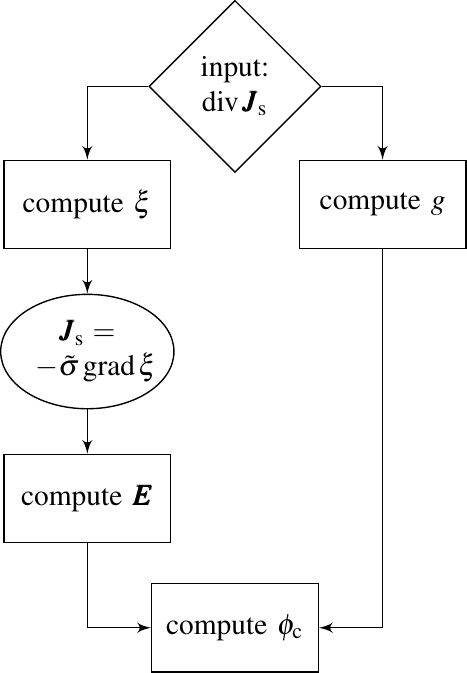}
	\caption{\label{fig:flow_chart} Steps for the computation of $\phi_\mathrm{c}$.}
\end{figure}

To calculate $\div\varepsilon_\mathrm{r} \bm{E}_\mathrm{s}$, the non-physical (since not charge conserving) situation $\bm{J} = \bm{J}_\mathrm{s}$ and $\rho = 0$ is considered. Here, the conducting structures of the DUT are not modeled as the conduction current density $\bm{J}_\mathrm{c}$ is disregarded. The correction term calculated from these sources therefore captures the isolated effects of the source current density $\bm{J}_\mathrm{s}$. By only considering source currents and no source charges, the ``source'' scalar potential $\phi_\mathrm{s}$ associated with  $\bm{E}_\mathrm{s}$ vanishes and the correction term only depends on the  ``source'' vector potential $\bm{A}_\mathrm{s}$,
\begin{equation}
\divv\varepsilon_\mathrm{r} \bm{E}_\mathrm{s}  = - \iu\omega\div\varepsilon_\mathrm{r} \bm{A}_\mathrm{s}
\end{equation}
A boundary value problem to calculate the scalar field $\div\varepsilon_\mathrm{r} \bm{A}_\mathrm{s}$ directly is found by applying the divergence operator to \eqref{eqn:potential_form_1}
\begin{equation}\label{eqn:div_As}
-\divv\varepsilon_\mathrm{r} \grad \div \varepsilon_\mathrm{r} \bm{A}_\mathrm{s} - \frac{\omega^2}{c^2}\div\varepsilon_\mathrm{r}\bm{A}_\mathrm{s} = \mu_0\div\bm{J}_\mathrm{s},
\end{equation}
and supplementing the same boundary conditions on the outer boundary $\partial \Omega$ as for the computation of $\xi$ in \eqref{eqn:grad_g}. The term $\div\bm{J}_\mathrm{s}$ is given with \eqref{eqn:div_J}.
To simplify the notation in the following, the scalar field
\begin{equation}
g := - \frac{1}{\mu_0} \div \varepsilon_\mathrm{r} \bm{A}_\mathrm{s}
\end{equation}
is defined. Expressing \eqref{eqn:div_As} with $g$ and supplying boundary conditions yields the BVP
\begin{subequations}\label{eqn:g_field}
	\begin{align}
	-\divv\varepsilon_\mathrm{r} \grad g - \frac{\omega^2}{c^2}g &=  -\div\bm{J}_\mathrm{s}  & & \mathrm{in} \; \Omega, \label{eqn:g_field_a} \\
	g &= \mathrm{const.} & & \mathrm{on} \; \Gamma_\mathrm{el}, \\
	\bm{n}\cdot\varepsilon_\mathrm{r}g &= 0 & & \mathrm{on} \; \Gamma_\mathrm{mag}.
	\end{align}
\end{subequations}
Using the definition of $g$ in \eqref{eqn:phi_2} and supplementing boundary conditions yields the BVP determining the compensated scalar potential $\phi_\mathrm{c}$,
\begin{subequations}\label{eqn:phi_3}
	\begin{align}
	-\divv\varepsilon_\mathrm{r}\grad\phi_\mathrm{c} -\frac{\omega^2}{c^2}\phi_\mathrm{c} &=  \div\varepsilon_\mathrm{r} \bm{E} +\iu\omega\mu_0g  & & \mathrm{in} \; \Omega, \\
	\phi_\mathrm{c} &= \mathrm{const.} & & \mathrm{on} \; \Gamma_\mathrm{el}, \\
	\bm{n}\cdot\varepsilon_\mathrm{r}\grad \phi_\mathrm{c} &= 0 & & \mathrm{on} \; \Gamma_\mathrm{mag}.
	\end{align}
\end{subequations}
This concludes the derivation of the field theoretical model. Thus, there are in total three steps to compute the compensated scalar potential $\phi_\mathrm{c}$ needed for the impedance calculation:
\begin{enumerate}
	\item Compute $\xi$ and $g$ from the $\divv \bm{J}_\mathrm{s}$ expression of \eqref{eqn:div_J} using \eqref{eqn:xi} and \eqref{eqn:g_field}, respectively.
	\item Compute $\bm{E}$ from $\bm{J}_\mathrm{s} = -\tilde{\sigma}\gradd \xi$ using \eqref{eqn:E_field_form}.
	\item Compute $\phi_\mathrm{c}$ from $\bm{E}$ and $g$ using \eqref{eqn:phi_3}.
\end{enumerate}
This procedure is illustrated in Fig~\ref{fig:flow_chart}.

The such computed fields are full-wave solutions considering all effects of Maxwell's equations, including retardation. Considering wave effects in a FE context generally requires a strategy preventing reflections from the outer boundary of the finite computational domain, usually employing perfectly matched layers (see, e.g., \cite{sacks_perfectly_1995}). However, if the structures of a DUT are small compared to the wavelengths associated with frequencies relevant to its EMC analysis, wave effects can generally be disregarded. To not unnecessarily complicate the FEM implementation and to avoid a possible greater numerical cost due to a finer space discretization needed for the layers, wave solutions can in such cases already be eliminated on the level of the underlying PDEs by employing the quasistatic Darwin approximation introduced in the following section \ref{sec:darwin}.

For applications that only require to consider inductive effects and ohmic losses the more restrictive magnetoquasistatic approximation introduced in section \ref{sec:mqs} is applicable, further reducing the complexity and numerical cost of the implementation.

\subsection{Darwin Approximation}
\label{sec:darwin}
Darwin's approximation \cite{larsson_electromagnetics_2007}, sometimes also referred to as ``full quasistatics'', neglects wave solutions while still fully capturing resistive, inductive, and capacitive effects. In this sense it behaves as the ``natural'' field-theoretical equivalent to the circuit model, which also precludes wave effects. The approximation arises from eliminating the magnetic vector potential contribution in both the displacement current and in Gauss' law, and is hence dependent on the gauge condition of the potentials. In the Lorenz gauge, it amounts to neglecting the $\omega^2$-terms in \eqref{eqn:potential_form}. Hence, also the $\omega^2$-term in \eqref{eqn:g_field_a} disappears in Darwin's approximation, yielding the frequency-independent BVP
\begin{subequations}\label{eqn:g_darwin}
	\begin{align}
	-\divv\varepsilon_\mathrm{r} \grad g &=  -\div\bm{J}_\mathrm{s}  & & \mathrm{in} \; \Omega, \\
	g &= \mathrm{const.} & & \mathrm{on} \; \Gamma_\mathrm{el}, \\
	\bm{n}\cdot\varepsilon_\mathrm{r}g &= 0 & & \mathrm{on} \; \Gamma_\mathrm{mag}.
	\end{align}
\end{subequations}
A comparison with \eqref{eqn:xi} determining the potential $\xi$ of the source current $\bm{J}_\mathrm{s}$ shows that in Darwin's approximation $g$ can be used to express the source current,
\begin{equation}\label{eqn:Js_darwin}
\bm{J}_\mathrm{s} = -\tilde{\sigma}\grad \xi = \varepsilon_\mathrm{r} \grad g.
\end{equation}
This is a great advantage if a model has to be evaluated at several frequency points, since in the Darwin case only one BVP must be solved to calculate both $\bm{J}_\mathrm{s}$ and $g$, in contrast to $1 + N_f$ BPVs in the non-approximated case, with $N_f$ being the number of frequency points to be evaluated.

As the vector potential contribution to the displacement current is neglected in the Darwin approximation, the E-field formulation changes to
\begin{equation}\label{eqn:pde_darwin}
\curll \nu_\mathrm{r} \curl \bm{E} + \iu\omega\mu_0\sigma \bm{E} + \frac{\omega^2}{c^2}\varepsilon_\mathrm{r}\grad\varphi = -\iu\omega\mu_0 \bm{J}_\mathrm{s}.
\end{equation}
The scalar potential $\varphi$ does not necessarily have to be the Lorenz-gauged scalar potential $\phi$ introduced in \eqref{eqn:gauge}. Since \eqref{eqn:pde_darwin} cannot be solved independently but only in a coupled BVP together with an additional scalar equation for $\varphi$, we choose $\varphi = \phi_\mathrm{c}$ such that \eqref{eqn:phi_3} can function as the required scalar equation. Thus, the coupled BVP reads
\begin{subequations} \label{eqn:BVP_darwin_general}
	\begin{align}
	\curll \nu_\mathrm{r} \curl \bm{E} +  \iu\omega\mu_0\sigma\bm{E} + \frac{\omega^2}{c^2}\varepsilon_\mathrm{r} \grad\phi_\mathrm{c} &&\notag\\
	= -\iu\omega\mu_0\varepsilon_\mathrm{r}\grad g && \mathrm{in} \; \Omega,\label{eqn:BVP_darwin_general_ampere}\\
	-\divv\varepsilon_\mathrm{r} \bm{E} - \divv\varepsilon_\mathrm{r}\grad\phi_\mathrm{c} -\frac{\omega^2}{c^2}\phi_\mathrm{c} &&\notag\\
	 = \iu\omega\mu_0g  && \mathrm{in} \; \Omega,\label{eqn:BVP_darwin_general_gauge}
	\end{align}
	\begin{align}
	\bm{n}\times\bm{E} = 0  \;\; &\mathrm{and\;\;} \phi_\mathrm{c} = \mathrm{const.} & &\mathrm{on} \; \Gamma_\mathrm{el}, \\
	\bm{n}\cdot\varepsilon_\mathrm{r}\bm{E} = 0 \;\;  &\mathrm{and\;\;} \bm{n}\cdot\varepsilon_\mathrm{r}\grad \phi_\mathrm{c} = 0 & &\mathrm{on} \; \Gamma_\mathrm{mag}.
	\end{align}
\end{subequations}

\subsection{Magnetoquasistatic Approximation}
\label{sec:mqs}
The magnetoquasistatic (MQS) approximation neglects the displacement current altogether and thereby also precludes capacitive effects in addition to wave effects. In contrast to Darwin's approximation, the $\omega^2$-terms of both \eqref{eqn:BVP_darwin_general_ampere} and \eqref{eqn:BVP_darwin_general_gauge} are neglected in the MQS formulation proposed here; while the first term is part of the displacement current, the second term results from the Lorenz gauge condition. Neglecting the latter is equivalent to choosing the Coulomb gauge condition for the calculation of $\phi$. This step is taken in the MQS approximation to establish a consistency at higher-frequencies between the two PDEs, which in numerical experiments proves to be necessary to obtain plausible results at arbitrarily high frequencies. The BVP to determine $\bm{E}$ in the MQS approximation is hence given by
\begin{subequations} \label{eqn:BVP_mqs1}
	\begin{align}
	\curll \nu_\mathrm{r} \curl \bm{E} &+  \iu\omega\mu_0\sigma\bm{E} &&\notag\\
	 &= -\iu\omega\mu_0\varepsilon_\mathrm{r}\grad g && \mathrm{in} \; \Omega,\\
	\bm{n}\times\bm{E} &= 0  && \mathrm{on} \; \Gamma_\mathrm{el}\\
	\bm{n}\cdot\varepsilon_\mathrm{r}\bm{E} &= 0  && \mathrm{on} \; \Gamma_\mathrm{mag}.
	\end{align}
\end{subequations}
The compensated scalar potential $\phi_\mathrm{c}$ is subsequently calculated with the BVP
\begin{subequations}\label{eqn:BVP_mqs2}
	\begin{align}
	-\divv\varepsilon_\mathrm{r}\grad\phi_\mathrm{c} &=  \div\varepsilon_\mathrm{r} \bm{E} +\iu\omega\mu_0g  & & \mathrm{in} \; \Omega, \\
		\phi_\mathrm{c} &= \mathrm{const.} & & \mathrm{on} \; \Gamma_\mathrm{el}, \\
		 \bm{n}\cdot\varepsilon_\mathrm{r}\grad \phi_\mathrm{c} &= 0 & & \mathrm{on} \; \Gamma_\mathrm{mag}.
	\end{align}
\end{subequations}
As in Darwin's approximation, for \eqref{eqn:BVP_mqs1} and \eqref{eqn:BVP_mqs2} the scalar field $g$ is determined with the frequency\hyp{}independent BVP \eqref{eqn:g_darwin}.

\subsection{Perfect Electric Conductor Approach}\label{sec:pec}
Due to the skin and proximity effects, the parasitics represented by the impedance matrix $\mathbf{Z}$ extracted using the general\hyp{}case equations of the previous subsections are dependent on the frequency in a general way. For some applications of EMC analysis, however, it may suffice to provide the high\hyp{}frequency limit which corresponds to a particular frequency dependence and to a decoupled pair of a frequency\hyp{}independent inductance matrix $\mathbf{L}$ and a frequency\hyp{}independent capacitance matrix $\mathbf{C}$. Approximating the parasitic effects with simple frequency\hyp{}independent lumped elements enables an especially straightforward combination with the functional elements in a joint circuit and a convenient and fast simulation thereof.

It is possible to extract the high-frequency inductance at a moderate numerical cost by modeling the conductors of the DUT as perfect electric conductors (PECs). This enforces a fully developed skin effect in the conductors such that the electric field $\bm{E}$ vanishes in the conducting domains. Then, at any frequency (below the first resonance) the same inductance matrix, i.e. its high-frequency limit, is obtained (Fig.~\ref{fig:inductance_1}).

The computational cost of this approach is much lower than for the general case for two reasons: First, since no frequency dependent behavior occurs, the BVPs have to be solved at only a single frequency point. Second, it is much cheaper to solve the E-field formulation \eqref{eqn:E_field_form} if the conductors are PECs since in this case \eqref{eqn:E_field_form_a} is only enforced in the non\hyp{}conducting region $\Omega_0 = \Omega \setminus \Omega_\mathrm{c}$, supplemented by electric boundary conditions for $\bm{E}$ on the boundary of the conducting region $\Omega_\mathrm{c}$. Furthermore, the ohmic loss term $\iu\omega\sigma\bm{E}$ disappears from the E-field form, such that the whole system of equations only needs to be solved for the imaginary part of the fields $\bm{E}$ and $\phi_\mathrm{c}$, since $\Re(\bm{E})=0$ and $\Re(\phi_\mathrm{c})=0$ if $\bm{J}_\mathrm{s}$ is chosen real. This leads to purely real operator matrices after FE discretization.

\subsubsection{PEC Case in Darwin's Approximation}
\label{sec:PEC_darwin}
In the PEC case, Darwin's approximation captures inductive and capacitive effects while precluding ohmic losses and wave effects. After solving \eqref{eqn:g_darwin} for $g$ in the full domain $\Omega$, the compensated potential $\phi_\mathrm{c}$ is computed with the BVP
\begin{subequations} \label{eqn:BVP_darwin_PEC}
	\begin{align}
	\curll \nu_\mathrm{r} \curl \bm{E} + \frac{\omega^2}{c^2}\varepsilon_\mathrm{r} \grad\phi_\mathrm{c}&&&\notag\\
	= -\iu\omega\mu_0\varepsilon_\mathrm{r}\grad g& & & \mathrm{in} \; \Omega_0,\label{eqn:BVP_darwin_PEC_a}\\
	-\divv\varepsilon_\mathrm{r} \bm{E} - \div\varepsilon_\mathrm{r}\grad\phi_\mathrm{c} -\frac{\omega^2}{c^2}\phi_\mathrm{c}&&&\notag\\
	 = \iu\omega\mu_0g& & &  \mathrm{in} \; \Omega, \label{eqn:BVP_darwin_PEC_b}
	\end{align}
	\begin{align}
	\bm{n}\times\bm{E} &= 0  & & \mathrm{on} \; \partial\Omega_\mathrm{c}\cup\Gamma_\mathrm{el},\\
	\phi_\mathrm{c} &= \mathrm{const.} & & \mathrm{on} \; \Gamma_\mathrm{el}, \\
	\bm{n}\cdot\varepsilon_\mathrm{r}\bm{E} = 0 \;\; & \mathrm{and} \;\; \bm{n}\cdot\varepsilon_\mathrm{r}\grad \phi_\mathrm{c} = 0 & & \mathrm{on} \; \Gamma_\mathrm{mag}.
	\end{align}
\end{subequations}
Note that while the vectorial equation \eqref{eqn:BVP_darwin_PEC_a} is only enforced in the non-conducting subdomain $\Omega_0$, the scalar equation \eqref{eqn:BVP_darwin_PEC_b} is enforced in the full domain $\Omega$.

\subsubsection{PEC Case in the MQS Approximation}
\label{sec:PEC_MQS}
Modeling the conductors as PECs in the MQS approximation yields a system that captures only inductive effects while precluding capacitive effects, ohmic losses, and wave effects. The corresponding BVP to determine $\bm{E}$ is given by
\begin{subequations} \label{eqn:BVP_mqs_pec}
	\begin{align}
	\curll \nu_\mathrm{r} \curl \bm{E} &= -\iu\omega\mu_0\varepsilon_\mathrm{r}\grad g & & \mathrm{in} \; \Omega_0,\\
	\bm{n}\times\bm{E} &= 0  & & \mathrm{on} \; \partial\Omega_\mathrm{c}\cup\Gamma_\mathrm{el}, \\
	\bm{n}\cdot\varepsilon_\mathrm{r}\bm{E} &= 0  & &  \mathrm{on} \; \Gamma_\mathrm{mag},
	\end{align}
\end{subequations}
It is enforced only in the non\hyp{}conducting subdomain $\Omega_0$. The BVPs determining $g$ and $\phi_\mathrm{c}$, \eqref{eqn:g_darwin} and \eqref{eqn:BVP_mqs2}, respectively, remain unchanged and are enforced in the full domain $\Omega$.

Capturing only inductive effects, the reactance calculated with \eqref{eqn:voltage_reduced_3d} and \eqref{eqn:impedance} from this $\phi_\mathrm{c}$ has a linear frequency dependence and its corresponding inductance is frequency independent (Fig.~\ref{fig:inductance_1}).  A real, frequency\hyp{}independent set of equations to determine the constant inductance of the MQS PEC case directly can therefore be obtained by dividing the BVPs \eqref{eqn:BVP_mqs_pec} and \eqref{eqn:BVP_mqs2}, and \eqref{eqn:voltage_reduced_3d} and \eqref{eqn:impedance} by $\iu\omega$ (introducing the scaled fields $\bm{E}' = \bm{E}/\iu\omega$ and $\phi_\mathrm{c}' = \phi_\mathrm{c}/\iu\omega$).

\subsection{Finite Element Discretization}
\label{sec:lfs}
We discretize the BVPs of the previous subsections by expressing the scalar fields $\xi$, $g$, and $\phi_\mathrm{c}$ with the $H^1$\hyp{}conforming and the vector field $\bm{E}$ with the $H(\mathrm{curl})$\hyp{}conforming basis functions given in \cite{ingelstrom_new_2006}. To this end, the domain $\Omega$ is meshed with tetrahedral elements. Testing in a Galerkin approach the scalar and vectorial PDEs with the scalar and vectorial basis functions, respectively, and integrating over the domain $\Omega$ discretizes the BVPs into sparse linear systems of equations.

FE discretizations of the frequency\hyp{}domain Maxwell equations are notorious for resulting in singular stiffness matrices at lower frequencies \cite{hiptmair_robust_2008,jochum_new_2015,eller_symmetric_2017}. Generally, a low\hyp{}frequency stabilization scheme is necessary to ensure that the associated FE method linear system has a stable solution at all frequencies. In \cite{eller_symmetric_2017} Eller \emph{et al.} derived a stable weak formulation based on the E-field formulation \eqref{eqn:E_field_form}. It is very general in its scope, such that it can be applied to the Darwin and MQS approximations, and to the PEC case in a straightforward way.

The basic approach of this method to split the Sobolev space $H(\mathrm{curl},\Omega)$ of the trial and test functions of $\bm{E}$ into three parts
\begin{equation}\label{eqn:sobolev_split}
	H(\mathrm{curl},\Omega) = V \oplus W \oplus U
\end{equation}
with
\begin{subequations}
	\begin{align}
	&\forall \bm{v} \in V\; \curl \bm{v} \neq 0,\label{eqn:spaces_1}\\
	&\forall \bm{w} \in W\; \curl \bm{w} = 0 \,\land\, \bm{w} \neq 0 \;\mathrm{in} \;\Omega_\mathrm{c},\label{eqn:spaces_2}\\
	&U := \{\bm{u} \in H(\mathrm{curl},\Omega) : \curl \bm{u} = 0 \, \land \, \bm{u} = 0 \;\mathrm{in} \;\Omega_\mathrm{c} \}.\label{eqn:spaces_3}
	\end{align}
\end{subequations}
Note that \eqref{eqn:spaces_3} defines $U$ whereas \eqref{eqn:spaces_1} and \eqref{eqn:spaces_2} are merely required conditions for the functions of the spaces $V$ and $W$, respectively. There are multiple ways to construct $V$ and $W$ such that \eqref{eqn:sobolev_split} is fulfilled.

The electric field is thereby decomposed into one part similar to a vector potential, $\bm{E}_V \in V$, and two parts that are essentially gradient fields produced by scalar potentials, $\bm{E}_W \in W$ and $\bm{E}_U \in U$, the latter of which vanishes in the conducting subdomain $\Omega_\mathrm{c}$. Moreover, the components are scaled differently, i.e.\
\begin{equation}
	\bm{E} = \iu\omega\bm{E}_V + (\iu\omega)^{1/2}\bm{E}_W + \bm{E}_U.
\end{equation}
Inserting this expression into the E-field formulation \eqref{eqn:E_field_form} and testing the equation separately with the functions of the three subspaces $V$, $W$, and $U$ allows an individual frequency scaling of the three resulting equations that renders the formulation low-frequency stable.
For edge-based FE basis functions of the lowest order $p=1$, the corresponding basis functions of $V$ are found using a tree-cotree split \cite{albanese_integral_1988}. The higher orders of the hierarchical basis functions of \cite{ingelstrom_new_2006} are already split into a part with non-vanishing curl and a gradient field part, and thereby give the higher-order functions of $V$ explicitly.

\section{Numerical Experiments}
\label{sec:experiments}
\subsection{Straight Wire}
The proposed FE-based method for impedance computation is evaluated by comparing its results to analytic values of the academic example of a straight wire of circular cross section. An analytic expression for the internal partial impedance $Z_\mathrm{int}$ of a wire of length $l$ and radius $r$ is given in \cite{ramo_fields_1994},
\begin{equation}
	Z_\mathrm{int} = R + \iu \omega L_\mathrm{int} = \frac{\iu l }{2\pi r}\sqrt{\frac{\omega\mu}{\sigma}}\left(\frac{\mathrm{Ber}(q)+\iu\mathrm{Bei}(q)}{\mathrm{Ber}'(q)+\iu\mathrm{Bei}'(q)}\right)
\end{equation}
with $q = r\sqrt{\omega\mu\sigma}$, and $\mathrm{Ber}$, $\mathrm{Bei}$, $\mathrm{Ber}'$ and $\mathrm{Bei}'$ denoting the real and imaginary Kelvin functions and their derivatives, respectively. This expression captures the ohmic resistance $R = \Re(Z)$ and the contribution of internal partial inductance $L_\mathrm{int}$ to the reactance $X = \Im(Z)$. It can be combined with the expression for the external partial inductance $L_\mathrm{ext}$ of a wire given in \cite{paul_inductance:_2010},
\begin{equation}
	L_\mathrm{ext} = \frac{\mu_0 l}{2\pi}\left(\mathrm{arsinh}\left(\frac{l}{r}\right)-\sqrt{1+\left(\frac{r}{l}\right)^2} + \frac{r}{l} \right),
\end{equation}
to produce an analytic approximation $Z_\mathrm{ana}$ that describes the ohmic resistance and the inductive contribution to the impedance of the wire,
\begin{equation}\label{eqn:Z_ana}
	Z_\mathrm{ana} := Z_\mathrm{int} + \iu\omega L_\mathrm{ext} = R + \iu\omega (L_\mathrm{int} + L_\mathrm{ext}).
\end{equation}
Neglecting both wave propagation and capacitive effects, $Z_\mathrm{ana}$ is an MQS approximation.
\begin{figure*}
	\begin{subfigure}{.5\textwidth}
		\includegraphics[scale=1]{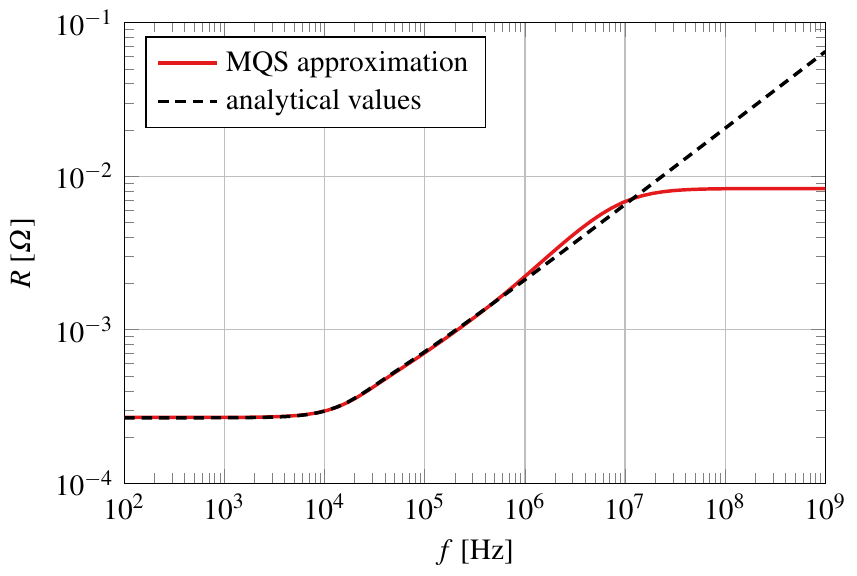}
		\caption{}
		\label{fig:R}
	\end{subfigure}
	\begin{subfigure}{.5\textwidth}
		\includegraphics[scale=1]{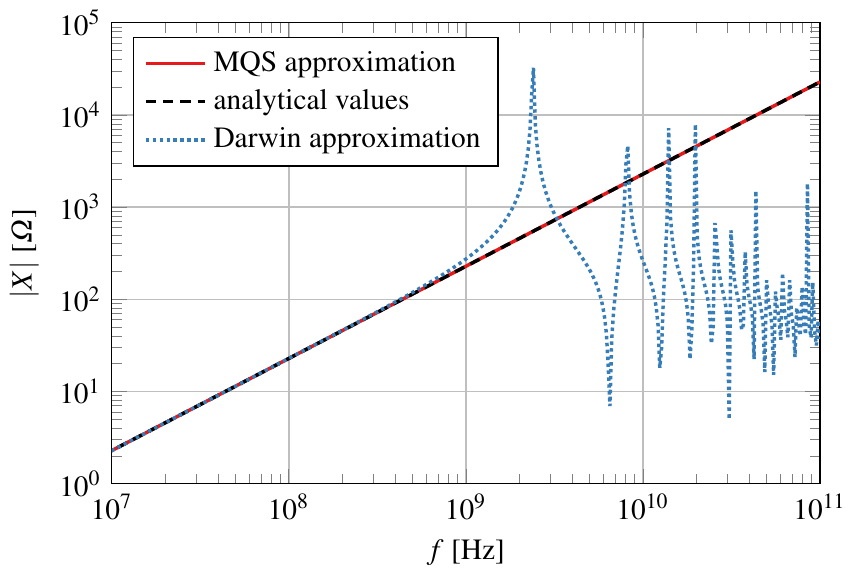}
		\caption{}
		\label{fig:X}
	\end{subfigure}
	\caption{Frequency dependent resistance $R$ (left) and modulus of the reactance $X$ (right). The FEM space discretization severely limits the applicability of the numerical method for high-frequency resistance computation.  While the Darwin approximation is capable of capturing resonant behavior in the reactance, the response of the MQS approximation is only inductive and hence linear.}
	\label{fig:RX}
\end{figure*}
\begin{figure*}
	\begin{subfigure}{.5\textwidth}
		\includegraphics[scale=1]{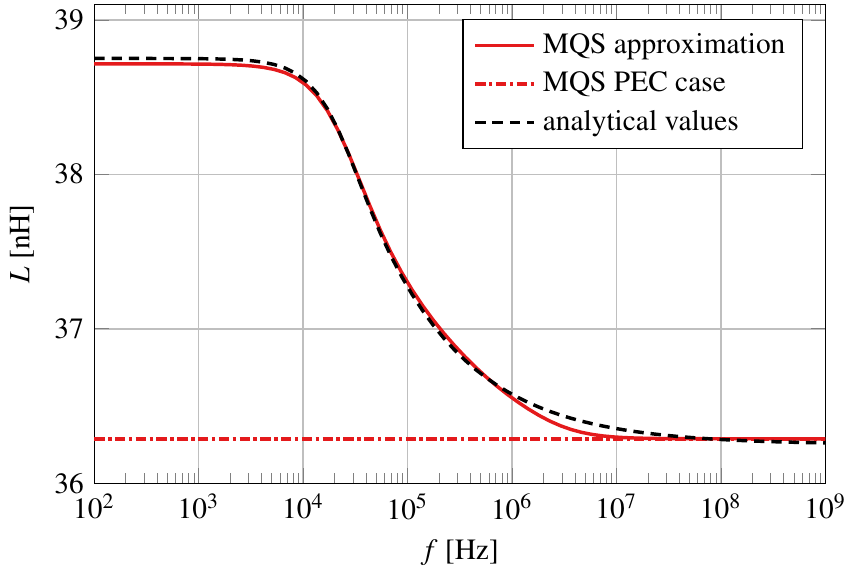}
		\caption{}
		\label{fig:inductance_1}
	\end{subfigure}
	\begin{subfigure}{.5\textwidth}
		\includegraphics[scale=1]{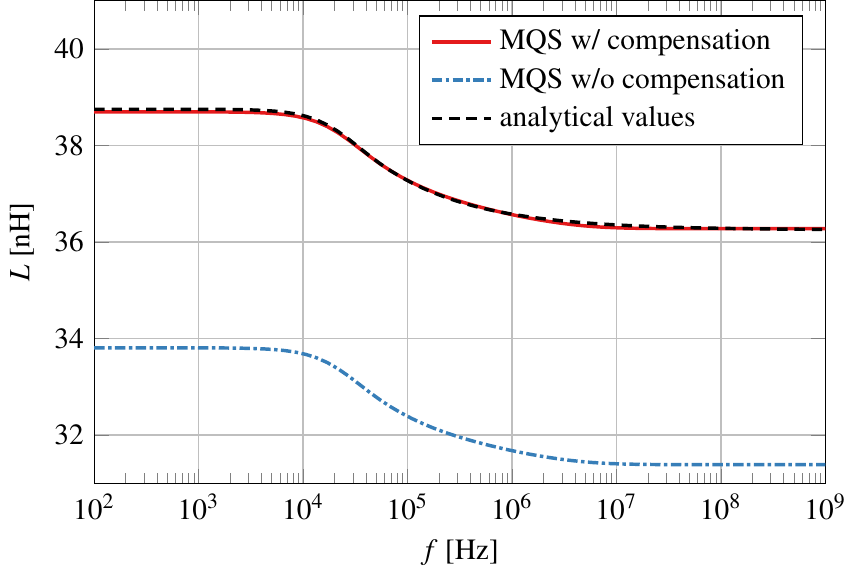}
		\caption{}
		\label{fig:inductance_2}
	\end{subfigure}
	\caption{Frequency dependence of the inductance $L$. The numerical values of the MQS approximation are in good agreement with the analytical values. The PEC case gives the high-frequency limit directly (left plot). The effect of the compensation term derived in section~\ref{sec:compensation} is demonstrated in the right plot.}
	\label{fig:inductance}
\end{figure*}

The example of a wire of length $l = \unit[50]{mm}$ and radius $r = \unit[1]{mm}$ was chosen for a series of numerical experiments. To approximate open boundary conditions in the FE implementation of the two quasistatic models of section~\ref{sec:bvp_general}, the impedance was calculated as the average of values produced with an electric boundary, $\partial \Omega = \Gamma_\mathrm{el}$, and values produced with a magnetic boundary, $\partial \Omega = \Gamma_\mathrm{mag}$, as suggested by the strategic dual image technique \cite{saito_finite_1987}, and the finite computational domain $\Omega$ was chosen very large compared to the model size.

Fig.~\ref{fig:R} compares the resistance over frequency calculated both analytically with \eqref{eqn:Z_ana} and numerically with the MQS approximation of section \ref{sec:mqs} (the $R$ values of the Darwin approximation and the non-approximated system are the same as the MQS values). The analytical and numerical values are virtually identical until $f=\unit[1]{MHz}$. For $f > \unit[10]{MHz}$, however, the numerical values quickly approach a constant upper limit due to the FE space discretization becoming unable to resolve the current density skin depth. This demonstrates the general infeasibility of numerical methods requiring a volumetric discretization for high-frequency resistance computations. In the context of EMC analysis, however, the resistive effects are generally only of minor importance. 

Fig.~\ref{fig:X} compares the modulus of the reactance $X$ of the analytical values and both quasistatic approximations in a higher frequency interval. While the analytical and numerical MQS values show the identical unchanging linear response, the Darwin approximation is capable of capturing resonant behavior resulting form the interplay of inductive and capacitive effects. This illustrates on the one hand that the MQS approximation is suitable for precisely those applications in which only the inductive effects are of interest, and on the other hand that the Darwin approximation may be employed for a resonance analysis (which has been investigated in \cite{traub_automated_2013}).

In the MQS approximation, the inductance $L$ is simply $L = X/\omega$. Fig.~\ref{fig:inductance} compares the inductance values of the numerical MQS approximation over frequency to the analytical values calculated from \eqref{eqn:Z_ana}. The values are in good agreement; unlike for the resistance $R$ in Fig.~\ref{fig:R} the finite minimal skin depth due to the space discretization does not lead to a major qualitative difference between the analytical and numerical values at higher frequencies. The finite discretization of the conductor in the numerical method manifests itself merely in a sightly altered curvature in the interval between $\unit[100]{kHz}$ and $\unit[10]{MHz}$. The values of the PEC case equation system \eqref{eqn:BVP_mqs_pec} displayed in Fig.~\ref{fig:inductance_1} show how this modification of the general case method produces the high-frequency limit of $L$ directly at any frequency. Fig.~\ref{fig:inductance_2} demonstrates the effect of the inductive compensation term derived in section \ref{sec:compensation} by comparing the values of the regular method with compensation to values calculated without the compensation term. The compensation leads to a frequency\hyp{}independent shift of the inductance of $\approx\unit[4.85]{nH}$. The uncompensated values are therefore on average $\approx \unit[13]{\%}$ lower than both the analytical and compensated numerical values.

\subsection{Capacitor Coil Model}
\begin{figure}
	\centering
	\includegraphics[width=0.4\textwidth]{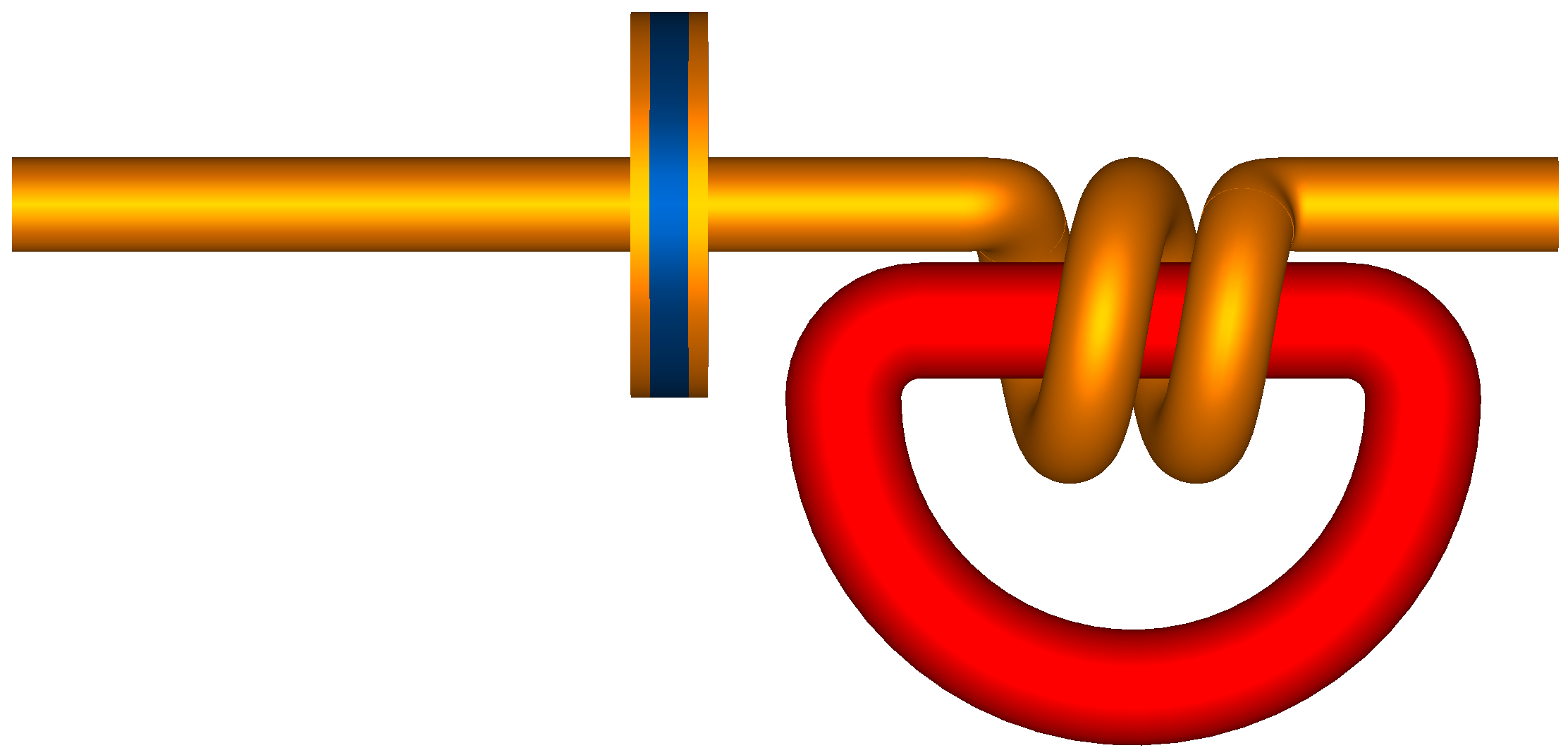}
	\caption{Model featuring a capacitor with a dielectric inset and a coil with a closed magnetic core. The total length of the model is $\unit[40]{mm}.$}
	\label{fig:LC_model}
\end{figure}

The capability of the proposed FE-based method for impedance computation to handle inhomogeneous permittivities $\varepsilon$ and permeabilities $\mu$ is demonstrated with a model featuring a parallel-plate capacitor with a dielectric inset, and a coil with a closed magnetic core, as displayed in Fig.~\ref{fig:LC_model}. Similar models without the dielectric inset and magnetic core have been considered by other authors, e.g.\ in \cite{hiptmair_robust_2008}.

Modeling the conductors of the model as PECs as described in section \ref{sec:pec} allows to extract a frequency\hyp{}independent capacitance $C$ for the capacitor and inductance $L$ for the coil. The capacitance $C$ is calculated with the Darwin system of section \ref{sec:PEC_darwin} by choosing the surfaces of the capacitor plates as terminal surfaces and computing the reactance $X_\mathrm{FE}$ between the two plates at a suitably low test frequency $f_0$. The capacitance is then given by $C = -1/(2\pi f_0 X_\mathrm{FE}(f_0))$. 

\begin{figure}
	\includegraphics[scale=1]{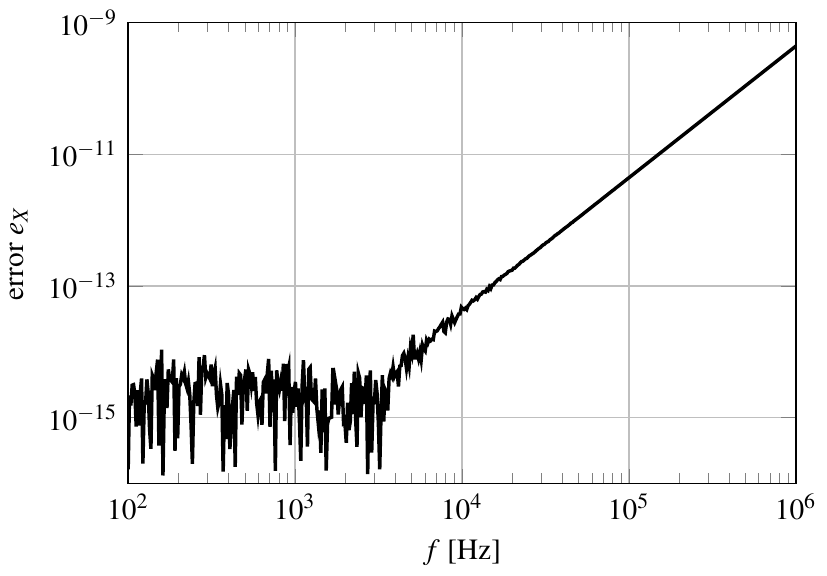}
	\caption{Frequency dependence of the relative error $e_X$ between the reactance $X_C$ of the extracted capacitance $C$ and the FE solution $X_\mathrm{FE}$, as defined in \eqref{eqn:error}.}
	\label{fig:LC_error}
\end{figure}

As an example for such a capacitance extraction, the model of Fig.~\ref{fig:LC_model} is considered with the choice $\varepsilon_\mathrm{r} = 100$ for the dielectric material and $\mu_\mathrm{r} = 1$ for the magnetic core. With the test frequency $f_0 = \unit[100]{Hz}$ the capacitance is computed as $C \approx \unit[1.184]{pF}$. The relative error
\begin{equation}\label{eqn:error}
e_X := \left|\frac{X_\mathrm{FM}-X_C}{X_\mathrm{FE}}\right|
\end{equation}
between the reactance $X_C := -1/(2\pi f C)$ associated with the extracted capacitance and the FE solution $X_\mathrm{FE}$ is displayed in Fig.~\ref{fig:LC_error}. The relative error $e_X$ stays below $10^{-14}$ for frequencies $f < \unit[5]{kHz}$, indicating both that in this frequency range the purely capacitive reactance $X_C$ perfectly captures the behavior of the FE solution $X_\mathrm{FE}$ and that the frequencies in this range are suitable test frequencies for the extraction of the capacitance $C$. Above $f=\unit[5]{kHz}$, the contribution of inductive effects to the reactance $X_\mathrm{FE}$ becomes measurable. However, the relative error $e_X$ stays well below $10^{-9}$ for frequencies as high as $f=\unit[1]{MHz}$.

\begin{figure*}
	\begin{subfigure}{.5\textwidth}
		\includegraphics[scale=1]{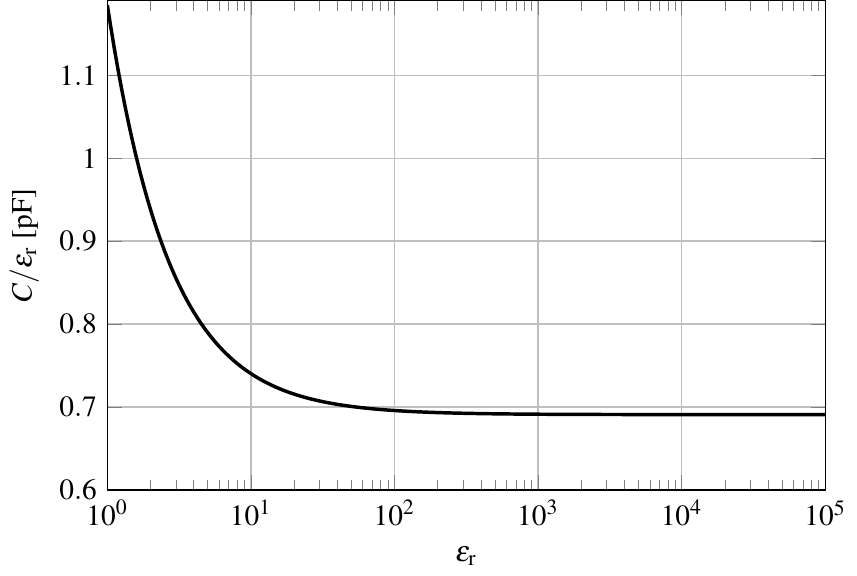}
		\caption{}
		\label{fig:LC_capacitance}
	\end{subfigure}
	\begin{subfigure}{.5\textwidth}
		\includegraphics[scale=1]{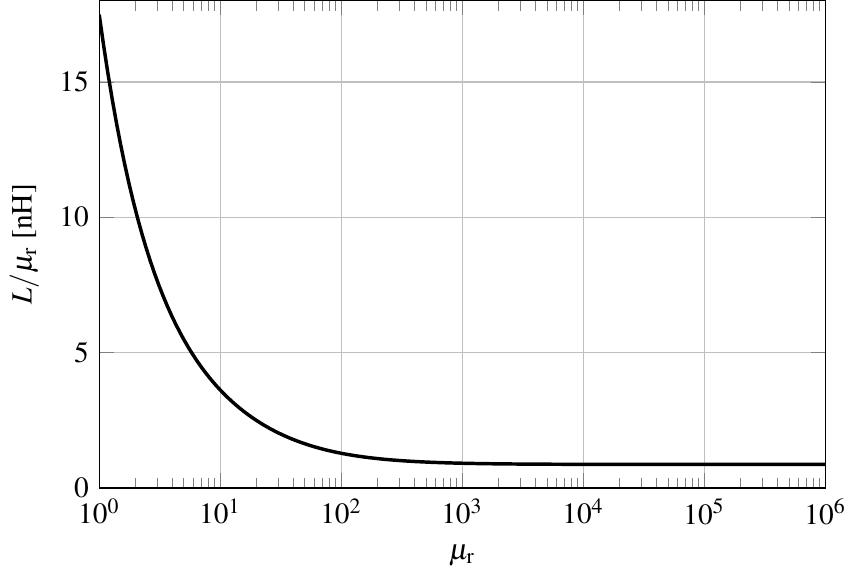}
		\caption{}
		\label{fig:LC_inductance}
	\end{subfigure}
	\caption{Normalized capacitance $C/\varepsilon_\mathrm{r}$ of the capacitor in dependence of the permittivity $\varepsilon_\mathrm{r}$ of the dielectric inset (left) and normalized inductance $L/\mu_\mathrm{r}$ of the coil vs.\ the permeability $\mu_\mathrm{r}$ of the core (right). Both quantities approach constant values for large values of $\varepsilon_\mathrm{r}$ and $\mu_\mathrm{r}$, respectively. This demonstrates the proportionality of $C$ and $L$ to their respective material parameters, that takes effect when the electric and magnetic fields get fully contained in the dielectric and core materials, respectively.}
\end{figure*}
Fig.~\ref{fig:LC_capacitance} shows the such extracted capacitance $C$ of the capacitor normalized by the relative permittivity $\varepsilon_\mathrm{r}$ of the dielectric material for a parameter sweep of $\varepsilon_\mathrm{r}$ (with the permeability of the magnetic core fixed at $\mu_\mathrm{r} = 1$). For higher values of $\varepsilon_\mathrm{r}$ the normalized capacitance  approaches a constant value $C/\varepsilon_\mathrm{r} \approx \unit[6.908]{pF}$. This demonstrates the proportionality of $C$ to relative permittivity  of the dielectric that takes effect when the field lines of the electric field $\bm{E}$ begin to be localized in the dielectric at values $\varepsilon_\mathrm{r} >1$.

Fig.~\ref{fig:LC_inductance} shows an analogous plot for the inductance $L$ of the coil normalized with the relative permeability $\mu_\mathrm{r}$ of the core (the terminal surfaces being the right capacitor plate and the right end surface of the coil, with the permittivity of the dielectric material fixed at $\varepsilon_\mathrm{r} =1$). The frequency\hyp{}independent inductance $L$ is computed directly with the MQS approximation as described in section \ref{sec:PEC_MQS}. The behavior of the normalized inductance $L/\mu_\mathrm{r}$ is qualitatively similar to that of $C/\varepsilon_\mathrm{r}$ in Fig.~\ref{fig:LC_capacitance}: For large values of $\mu_\mathrm{r}$ the normalized inductance approaches a constant value $L/\mu_\mathrm{r} \approx \unit[872.1]{pH}$, indicating the proportionality of $L$ to $\mu_r$ for values $\mu_r \gg 1$, at which the field lines of the magnetic flux $\bm{B}$ are concentrated in the core.

\section{Conclusions}
\label{sec:conclusion}
In his work, the theoretical foundation of an FE-based numerical impedance computation method was developed, providing a basis for parasitic extraction for models of complex geometry. After clarifying the relationship of the impedance matrix and a scalar potential, the field equations for the calculation of the scalar potential were formulated and a suitable source current density was modeled. A de-embedding approach was pursued to eliminate any unwanted inductive influence of the current source on the result. The FE discretization including a necessary low-frequency stabilization scheme was briefly discussed. While the derivation itself did not rely on any approximation to Maxwell's equations, two quasistatic approximations especially relevant for application were also provided, as well as the important special case of perfect electric conductors giving a high-frequency inductance approximation. The method was validated by comparing the quasistatic approximations and the special PEC case to analytic values for a wire model. The results demonstrate that the finite minimal skin depth of the FE discretization does not negatively influence the quality of high\hyp{}frequency inductance results in a substantial way. The capability of the method to handle inhomogeneous permittivities and permeabilities was demonstrated with a capacitor coil model featuring a dielectric inset and a magnetic core.


%
%

\bibliographystyle{spbasic_unsrt}      
\bibliography{literature}{}

\begin{thebibliography}{20}
\providecommand{\natexlab}[1]{#1}
\providecommand{\url}[1]{{#1}}
\providecommand{\urlprefix}{URL }
\expandafter\ifx\csname urlstyle\endcsname\relax
  \providecommand{\doi}[1]{DOI~\discretionary{}{}{}#1}\else
  \providecommand{\doi}{DOI~\discretionary{}{}{}\begingroup
  \urlstyle{rm}\Url}\fi
\providecommand{\eprint}[2][]{\url{#2}}

\bibitem[{Ruehli(1974)}]{ruehli_equivalent_1974}
Ruehli AE (1974) Equivalent circuit models for three-dimensional multiconductor
  systems. IEEE Transactions on Microwave Theory and Techniques 22(3):216--221

\bibitem[{Kamon et~al.(1998)Kamon, Marques, Silveira, and
  White}]{kamon_automatic_1998}
Kamon M, Marques NA, Silveira LM, White J (1998) Automatic generation of
  accurate circuit models of 3-{D} interconnect. IEEE Transactions on
  Components, Packaging, and Manufacturing Technology: Part B 21(3):225--240

\bibitem[{Kamon et~al.(1994)Kamon, Ttsuk, and White}]{kamon_fasthenry:_1994}
Kamon M, Ttsuk M, White J (1994) {FASTHENRY}: a multipole-accelerated 3-{D}
  inductance extraction program. IEEE Transactions on Microwave Theory and
  Techniques 42(9):1750--1758, \doi{10.1109/22.310584}

\bibitem[{Ruehli et~al.(2003)Ruehli, Antonini, Esch, Ekman, Mayo, and
  Orlandi}]{ruehli_nonorthogonal_2003}
Ruehli AE, Antonini G, Esch J, Ekman J, Mayo A, Orlandi A (2003) Nonorthogonal
  {PEEC} formulation for time- and frequency-domain {EM} and circuit modeling.
  IEEE Transactions on Electromagnetic Compatibility 45(2):167--176,
  \doi{10.1109/TEMC.2003.810804}

\bibitem[{Traub et~al.(2012)Traub, Hansen, Ackermann, and
  Weiland}]{traub_generation_2012}
Traub F, Hansen J, Ackermann W, Weiland T (2012) Generation of physical
  equivalent circuits using 3d simulations. In: 2012 {IEEE} {International}
  {Symposium} on {Electromagnetic} {Compatibility}, IEEE, pp 486--491

\bibitem[{Schuhmacher et~al.(2018)Schuhmacher, Klaedtke, Keller, Ackermann, and
  De~Gersem}]{schuhmacher_adjoint_2018}
Schuhmacher S, Klaedtke A, Keller C, Ackermann W, De~Gersem H (2018) Adjoint
  {Technique} for {Sensitivity} {Analysis} of {Coupling} {Factors} {According}
  to {Geometric} {Variations}. IEEE Transactions on Magnetics 54(3):1--4,
  \doi{10.1109/TMAG.2017.2774107}

\bibitem[{Jordan and Balmain(1968)}]{jordan_electromagnetic_1968}
Jordan EC, Balmain KG (1968) Electromagnetic waves and radiating systems, 2nd
  edn. Prentice-{Hall} electrical engineering series, Prentice-Hall, Englewood
  Cliffs, NJ, oCLC: 439397

\bibitem[{Paul(2010)}]{paul_inductance:_2010}
Paul CR (2010) Inductance: {Loop} and {Partial}. Wiley; IEEE, Hoboken, N.J.,
  oCLC: ocn428031806

\bibitem[{Monk(2003)}]{monk_finite_2003}
Monk P (2003) Finite element methods for {Maxwell}'s equations. Numerical
  mathematics and scientific computation, Clarendon Press ; Oxford University
  Press, Oxford : New York, oCLC: ocm51109019

\bibitem[{Nisbet and Kemmer(1957)}]{nisbet_electromagnetic_1957}
Nisbet A, Kemmer N (1957) Electromagnetic potentials in a heterogeneous
  non-conducting medium. Proceedings of the Royal Society of London Series A
  Mathematical and Physical Sciences 240(1222):375--381,
  \doi{10.1098/rspa.1957.0092}

\bibitem[{Sacks et~al.(1995)Sacks, Kingsland, Lee, and
  Lee}]{sacks_perfectly_1995}
Sacks ZS, Kingsland DM, Lee R, Lee JF (1995) A perfectly matched anisotropic
  absorber for use as an absorbing boundary condition. IEEE Transactions on
  Antennas and Propagation 43(12):1460--1463

\bibitem[{Larsson(2007)}]{larsson_electromagnetics_2007}
Larsson J (2007) Electromagnetics from a quasistatic perspective. American
  Journal of Physics 75(3):230--239, \doi{10.1119/1.2397095}

\bibitem[{Ingelström(2006)}]{ingelstrom_new_2006}
Ingelström P (2006) A new set of {H}(curl)-conforming hierarchical basis
  functions for tetrahedral meshes. IEEE Transactions on Microwave Theory and
  Techniques 54(1):106--114, \doi{10.1109/TMTT.2005.860295}

\bibitem[{Hiptmair et~al.(2008)Hiptmair, Kramer, and
  Ostrowski}]{hiptmair_robust_2008}
Hiptmair R, Kramer F, Ostrowski J (2008) A {Robust} {Maxwell} {Formulation} for
  {All} {Frequencies}. IEEE Transactions on Magnetics 44(6):682--685,
  \doi{10.1109/TMAG.2007.915991}

\bibitem[{Jochum et~al.(2015)Jochum, Farle, and
  Dyczij-Edlinger}]{jochum_new_2015}
Jochum M, Farle O, Dyczij-Edlinger R (2015) A {New} {Low}-{Frequency} {Stable}
  {Potential} {Formulation} for the {Finite}-{Element} {Simulation} of
  {Electromagnetic} {Fields}. IEEE Transactions on Magnetics 51(3):1--4,
  \doi{10.1109/TMAG.2014.2360080}

\bibitem[{Eller et~al.(2017)Eller, Reitzinger, Schöps, and
  Zaglmayr}]{eller_symmetric_2017}
Eller M, Reitzinger S, Schöps S, Zaglmayr S (2017) A {Symmetric}
  {Low}-{Frequency} {Stable} {Broadband} {Maxwell} {Formulation} for
  {Industrial} {Applications}. SIAM Journal on Scientific Computing
  39(4):B703--B731, \doi{10.1137/16M1077817}

\bibitem[{Albanese and Rubinacci(1988)}]{albanese_integral_1988}
Albanese R, Rubinacci G (1988) Integral formulation for 3d eddy-current
  computation using edge elements. IEE Proceedings A (Physical Science,
  Measurement and Instrumentation, Management and Education, Reviews)
  135(7):457--462

\bibitem[{Ramo et~al.(1994)Ramo, Whinnery, and Van~Duzer}]{ramo_fields_1994}
Ramo S, Whinnery JR, Van~Duzer T (1994) Fields and waves in communication
  electronics, 3rd edn. Wiley, New York

\bibitem[{Saito et~al.(1987)Saito, Takahashi, and Hayano}]{saito_finite_1987}
Saito Y, Takahashi K, Hayano S (1987) Finite element solution of open boundary
  magnetic field problems. IEEE Transactions on Magnetics 23(5):3569--3571,
  \doi{10.1109/TMAG.1987.1065581}

\bibitem[{Traub et~al.(2013)Traub, Hansen, Ackermann, and
  Weiland}]{traub_automated_2013}
Traub F, Hansen J, Ackermann W, Weiland T (2013) Automated construction of
  physical equivalent circuits for inductive components. In: 2013
  {International} {Symposium} on {Electromagnetic} {Compatibility}, IEEE, pp
  67--72

\end{thebibliography}

\end{document}